\documentclass[aps,pre,groupedaddress,reprint,showpacs,pdftex]{revtex4-1}
\usepackage{setspace}
\usepackage{here}
\usepackage{url}
\usepackage[symbol]{footmisc}
\usepackage{graphicx}
\usepackage{color}

\newcommand{\eqref}{\ref}
\setcitestyle{super}
\begin{document}
\title{
  Making the most of data:
  Quantum Monte Carlo Post-Analysis Revisited
}
\author{Tom Ichibha}
\affiliation{School of Information Science, JAIST, 1-1 Asahidai, Nomi, Ishikawa, 923-1292, Japan.}
\author{Kenta Hongo}
\affiliation{Research Center for Advanced Computing Infrastructure, JAIST, 1-1 Asahidai, Nomi, Ishikawa 923-1292, Japan.}
\affiliation{Center for Materials Research by Information Integration, Research and Services Division of Materials Data and Integrated System, National Institute for Materials Science, 1-2-1 Sengen, Tsukuba 305-0047, Japan.}
\affiliation{PRESTO, Japan Science and Technology Agency, 4-1-8 Honcho, Kawaguchi-shi, Saitama 322-0012, Japan.}
\affiliation{Computational Engineering Applications Unit, RIKEN, 2-1 Hirosawa, Wako, Saitama 351-0198, Japan.}
\author{Ryo Maezono}
\affiliation{School of Information Science, JAIST, 1-1 Asahidai, Nomi, Ishikawa, 923-1292, Japan.}
\affiliation{Computational Engineering Applications Unit, RIKEN, 2-1 Hirosawa, Wako, Saitama 351-0198, Japan.}
\author{Alex J.W. Thom}
\email{ajwt3@cam.ac.uk}
\affiliation{Department of Chemistry, University of Cambridge, Lensfield Road, Cambridge, CB2 1EW, U.K.}
\newpage

\begin{abstract}
  In quantum Monte Carlo (QMC) methods, energy estimators are calculated
  as the statistical average of the Markov chain sampling
  of energy estimator along with an associated statistical error.
  This error estimation is not straightforward and
  there are several choices of the error estimation methods.
  We evaluate the performance of three methods, Straatsma,
  an autoregressive model, and
    a blocking analysis based on von Neumann's ratio test for randomness,
  for the energy
  time-series given by Diffusion Monte Carlo, Full Configuration Interaction
  Quantum Monte Carlo and Coupled Cluster Monte Carlo methods. From these analyses
  we describe a hybrid analysis method which provides reliable error estimates for
  series of all lengths.
  Equally important is the estimation of the appropriate start point of
  the equilibrated phase, and two heuristic schemes are tested, establishing
  that MSER (mean squared error rule) gives reasonable and constant estimations
  independent of the length of time-series.
\end{abstract}
\date{\today}
\maketitle

\section{Introduction}
\label{sec.intro}
With the increase in availability of large-scale computers, Quantum Monte Carlo (QMC) methods
have spread rapidly owing to the embarassing parallelizability of such algorithms. \cite{2018SPE,2015BLU,2017MAT} 
QMC is one of the most accurate {\it ab initio} methods,
and it is often used for systems which cannot be sufficiently accurately 
described by Density Functional Theory (DFT) \cite{2017ICH,2016LUO,2017TRA,2018KEN}
or which are too large to apply post Hartree--Fock methods.\cite{2018BEN}

\vspace{2mm}
Diffusion Monte Carlo (DMC)\cite{2001FOU} is one such 
QMC method with a computational scaling of $\mathcal{O}(N^3)$ for a system of $N$ electrons,
and, as such, it can be applied even to large-sized systems
including more than 1000 electrons. \cite{2016LUO}
The drawback of DMC is the requirement to use the fixed-node approximation\cite{2006CAN}
to avoid the sign problem, which introduces a systematic error dependent upon the quality
of the nodes of a trial wavefunction, and, although there are ways to suppress
this error,\cite{2006RIO,2012MOR} they make calculations considerably more expensive. 

\vspace{2mm}
Two newer QMC methods in quantum chemistry have attracted interest of late, as they are
not constrained by the fixed-node approximation:
Firstly the full-configuration interaction QMC (FCIQMC) method,\cite{2010BOO, 2010CLE, 2011CLE}
which stochastically solves the equations of full-configuration interaction (FCI), by sampling with discrete particles.
Although the scaling of calculation cost is still exponential\cite{2011CLE} in the number of electrons
like FCI, the prefactor of scaling curve is significantly reduced.
Thus, this method can be applied to medium-sized systems.\cite{2017SCO}
Secondly is Coupled Cluster Monte Carlo (CCMC), which stochastically solves
Coupled Cluster (CC) equations.\cite{2010THO_ccmc, 2016SPE}
Since the parameter space of a truncated CC calculation is smaller than that
of FCI, CCMC will in general have a smaller memory cost than FCIQMC.

\vspace{2mm}
QMC methods commonly provide an energy estimator 
as the statistical average of sampling a Markov chain,
also producing an estimate of the statistical error. 
It is difficult to estimate the error reliably
due to the following reasons:\cite{2017ALE}
(i)~The samples are not independent of each other
but correlated along the simulation time evolution.
(ii)~When the distribution of sampling is non-normal,
the probability distribution of the mean value
is also non-normal unless the number of sampling
is large enough to satisfy the central limit theorem.
In this work, we examine the performance of three
characteristic automatic error estimation methods,
Straatsma, \cite{1986STR}, the AutoRegressive
(AR) model \cite{2010THO_ar}, and 
  blocking analysis based on
  von Neumann's ratio test for randomness \cite{1978FIS,2011YOU}
  (von Neumann blocking)
for the energy time-series obtained by applying DMC, FCIQMC,
and CCMC to the neon atom.
From these data we establish recommendations for the most reliable error
estimation method for different lengths of time series, and devise
a new hybrid scheme applicable to any length of time series.

\vspace{2mm}        
Another important issue on the post-analysis of QMC is
to determine the length of the pre-equilibration (warm-up) phase.\cite{2017ALE}
Underestimation of the length gives a systematic error in the energy
but its overestimation also increases the statistical error.
In this work, we tested two heuristic methods to estimate the warm-up steps. 
One is MSER (mean squared error rule)~\cite{2008FRA} and the other is
min-WREE (minimization of weighted relative error of the error) inspired 
by the post-analysis implemented in HANDE QMC code for stochastic
quantum chemistry. \cite{2014SPE,2019SPE} 
Our analysis establishes that the estimation of warm-up steps 
by min-WREE changes depending on the length of time-series.  
On the other hand, MSER makes reasonable and constant estimation of
warm-up steps, independent of the length of time-series.

\begin{figure}[htbp]
  \centering
  \includegraphics[width=1.0\hsize]{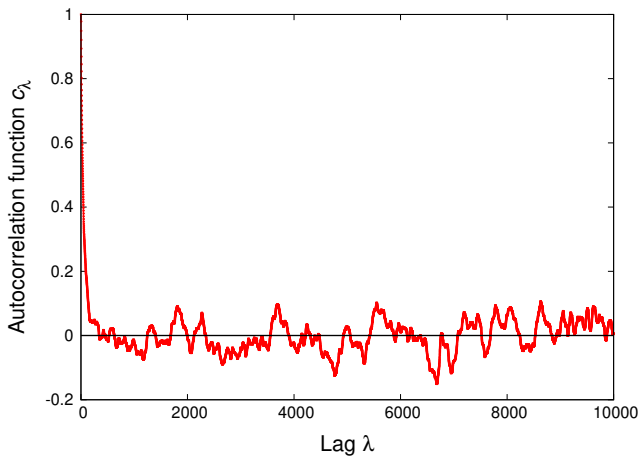}
  \caption{
    A typical autocorrelation function of energy time-series
    generated by CCMC calculation.
    It rapidly decreases and oscillates around zero as the lag,
    $\lambda$, increases, showing that the correlation
    between $X_i$ and $X_{i+\lambda}$ decreases.   
  }
  \label{fig.autocorr}
\end{figure}

\section{Error estimation method}
\label{sec.methods}        
In this section we elucidate the error estimation methods used:
Straatsma,\cite{1986STR} AR model,\cite{2010THO_ar} and von Neumann
blocking.\cite{2011YOU}

\subsection*{Straatsma}
\label{subsec.straatsma}
Straatsma \textit{et al.} show that the variance of
the statistical average of stationary time-series including
$n$ samples $\left\{ {{X_i}} \right\}_{i = 1}^n$ is given as
follows without any assumptions:\cite{1986STR}
\begin{eqnarray}
  \sigma _{\bar X}^2 &=& \frac{{{c_0}}}{n}\left[ {1 + 2\sum\limits_{\lambda  = 1}^{n - 1} {\left( {1 - \frac{\lambda }{n}} \right){c_\lambda }} } \right] = {c_0}\frac{\tau }{n},
  \label{eq.stra1}
  \\
  \tau  &\equiv& 1 + 2\sum\limits_{\lambda  = 1}^{n - 1} {\left( {1 - \frac{\lambda }{n}} \right){c_\lambda }}.
  \label{eq.stra3}
\end{eqnarray}
Here, $\tau$ is the estimation of time-correlation length (steps).%
\footnote[0]{
  The correlation length $\tau$ is differently defined in two papers,
  \cite{1986STR, 2010THO_ar} but they give the same definition of error.
  We took the newer definition of $\tau$.\cite{2010THO_ar}
}
${{c_\lambda }}$ is the autocorrelation function with lag $\lambda$,
which is approximately given by the finite number $n$ of samples as:
\begin{eqnarray}
  {c_\lambda } \approx \frac{1}{{n - \lambda }}\sum\limits_{i = 1}^{n-\lambda} {\left\{ {\left( {{X_i} - \bar X} \right)\left( {{X_{i + \lambda }} - \bar X} \right)} \right\}},\;\;(\lambda<n).
  \label{eq.stra2}  
\end{eqnarray}
Here, the approximation of $c_\lambda$ is inaccurate
when the number of terms $n - \lambda$ to be summed up is small.
Thus, in equation \eqref{eq.stra3} we limit the summation over $\lambda$ to values before $c_\lambda$ becomes negative
for the first time, since later $c_\lambda$ oscillates around zero
as shown in Figure \ref{fig.autocorr}.  The resulting $\tau$ is then used to calculate   $\sigma _{\bar X}^2$.


\subsection*{AutoRegressive (AR) Model}
The AR model assumes that the random process of
Markov chain sampling $\left\{ {{X_i}} \right\}_{i = 1}^n$
can be reasonably described by  \cite{2010THO_ar}
\begin{eqnarray}
  {X_i} &=& \bar X + {\pi _1}{X_{i - 1}} + {\pi _2}{X_{i - 2}}
  +  \cdots  + {\pi _p}{X_{i - p}} + {a_i}, 
  \label{eq.ar1}\\
  \bar X &=& \frac{1}{n}\left( {{X_1} + {X_2} +  \cdots  + {X_n}} \right)
\end{eqnarray}
The $i$-th sample is given as a linear combination of
the previous steps and a random Gaussian noise $a_i$
with average 0 and variance $\sigma_a^2$. The coefficients
$\left\{ {{\pi _i}} \right\}_{i = 1}^p$ and the variance
$\sigma_a^2$ are fitted to the given time-series
using Yule--Walker equation.\cite{2006WEI}
The number of coefficients $p$ is decided based on
Akaike's Information Criterion (AIC).~\cite{1992AKA}
  Large number of coefficients are needed to accurately
  describe the stochastic process of the given time-series,
  but, if it is too large, it becomes over-fitting.
  AIC aims to provide an appropriate compromise $p$ value.

The estimation of the correlation length $\tau$ is calculated by
\begin{eqnarray}
  \tau  = {{\left( {1 - \sum\limits_{\lambda  = 1}^p {{c_\lambda }{\pi _\lambda }} } \right)} \mathord{\left/
 {\vphantom {{\left( {1 - \sum\limits_{\lambda  = 1}^p {{c_\lambda }{\pi _\lambda }} } \right)} {{{\left( {1 - \sum\limits_{\lambda  = 1}^p {{\pi _\lambda }} } \right)}^2}}}} \right.
 \kern-\nulldelimiterspace} {{{\left( {1 - \sum\limits_{\lambda  = 1}^p {{\pi _\lambda }} } \right)}^2}}},
  \label{eq.ar2}
\end{eqnarray}
where $c_\lambda$ are defined by equation \eqref{eq.stra2}.

\subsection*{von Neumann blocking}
Von Neumann blocking takes into account the non-normality
of the distribution of the sampling, in contrast to 
the two above-mentioned methods. The given time-series
$\left\{ {{X_i}} \right\}_{i = 1}^n$ is divided into
blocks with block size $m$, and a new time-series produced:
\begin{eqnarray}
  {W_j}(m) = \frac{1}{m}\sum\limits_{l = 1}^m {{X_{m\left( {j - 1} \right) + l}}}.
\end{eqnarray}
Both the non-normality of the distribution and the correlation
length are reduced in the new time-series
$\left\{ {{W_j}} \right\}_{i = 1}^n$.
The block size $m$ is decided such that each ${W_j}(m)$ can be regarded 
to be sampled from independent and identical normal distributions,
based on von Neumann's rate test for randomness.\cite{1978FIS}
After blocking, the variance of the statistical average
is given by
$\sigma _{\bar X}^2 = \frac{1}{{k\left( {k - 1} \right)}}\sum\limits_{j = 1}^k {{{\left[ {{W_j}\left( m \right) - \bar W} \right]}^2}}$,~$\left( {\bar W \equiv \bar X} \right)$.

\vspace{2mm}
  Here, we also note the very commonly used blocking approach by
  Flyvbjerg and Petersen.\cite{1989FLY}
  This method performs integration of autocorrelation functions
  almost equivalent to equation \eqref{eq.stra1} yet utilizing
  blocking, aiming to reduce the computational cost of analysis:
  It is mathematically similar to Straatsma \cite{1986STR}
  but it suffers a systematic bias stemming from blocking
  as well as von Neumann blocking.
\begin{figure*}[htbp]
  \begin{minipage}{0.4\hsize}
    \centering
    \includegraphics[width=\hsize]{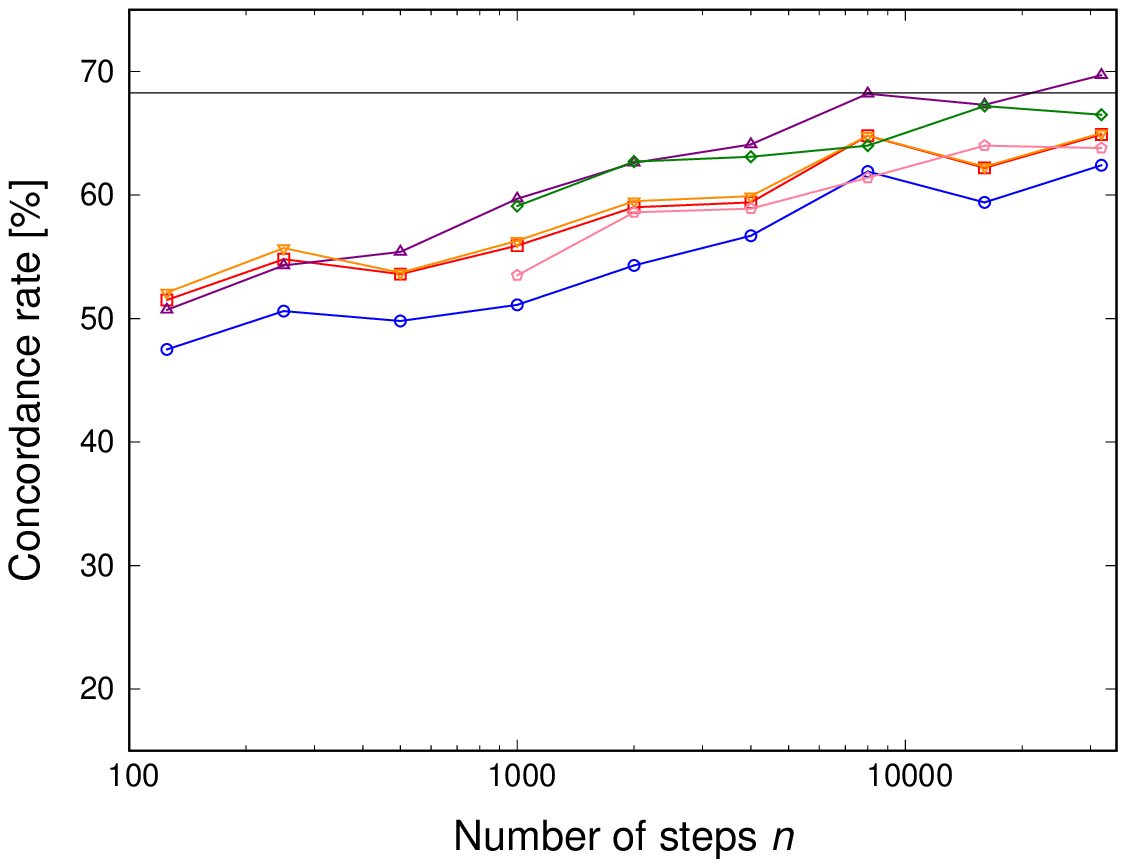}
    (a)~DMC
  \end{minipage}
  \begin{minipage}{0.4\hsize}
    \centering
    \includegraphics[width=\hsize]{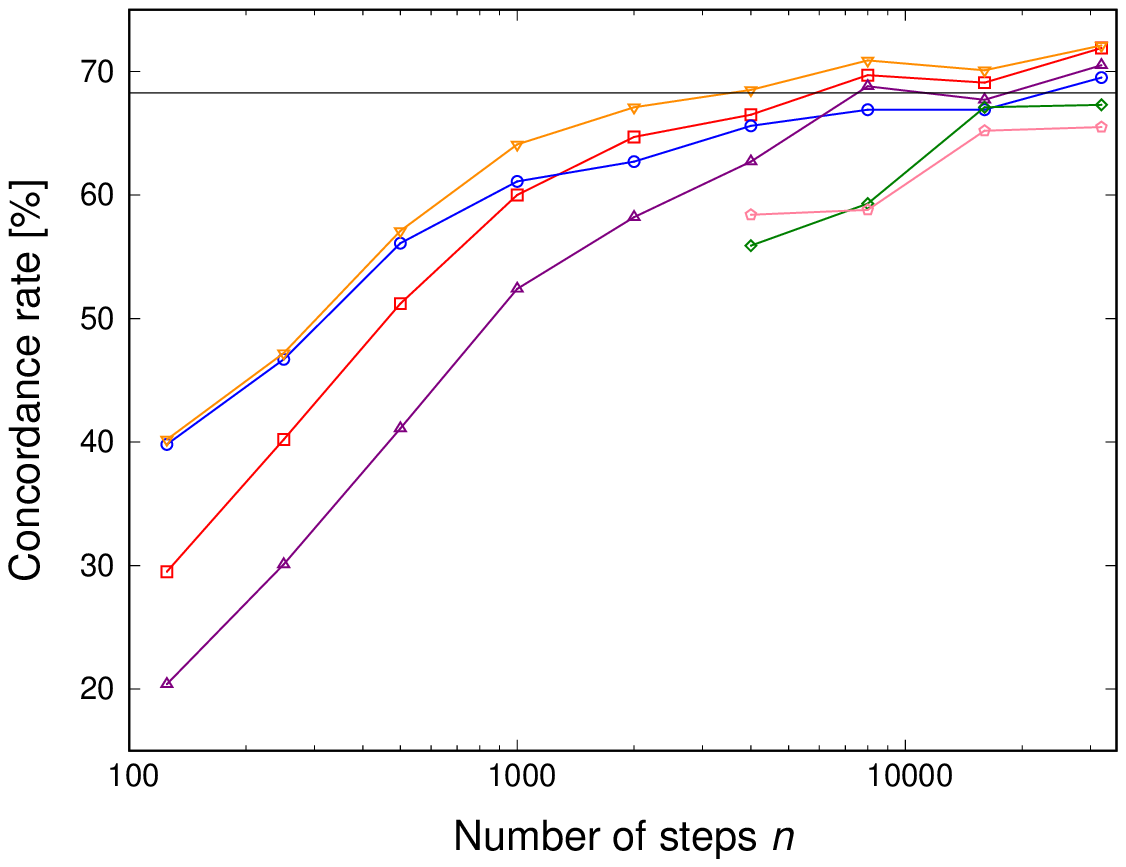}
    (b)~FCIQMC
  \end{minipage}
  \begin{minipage}{0.4\hsize}
    \centering
    \includegraphics[width=\hsize]{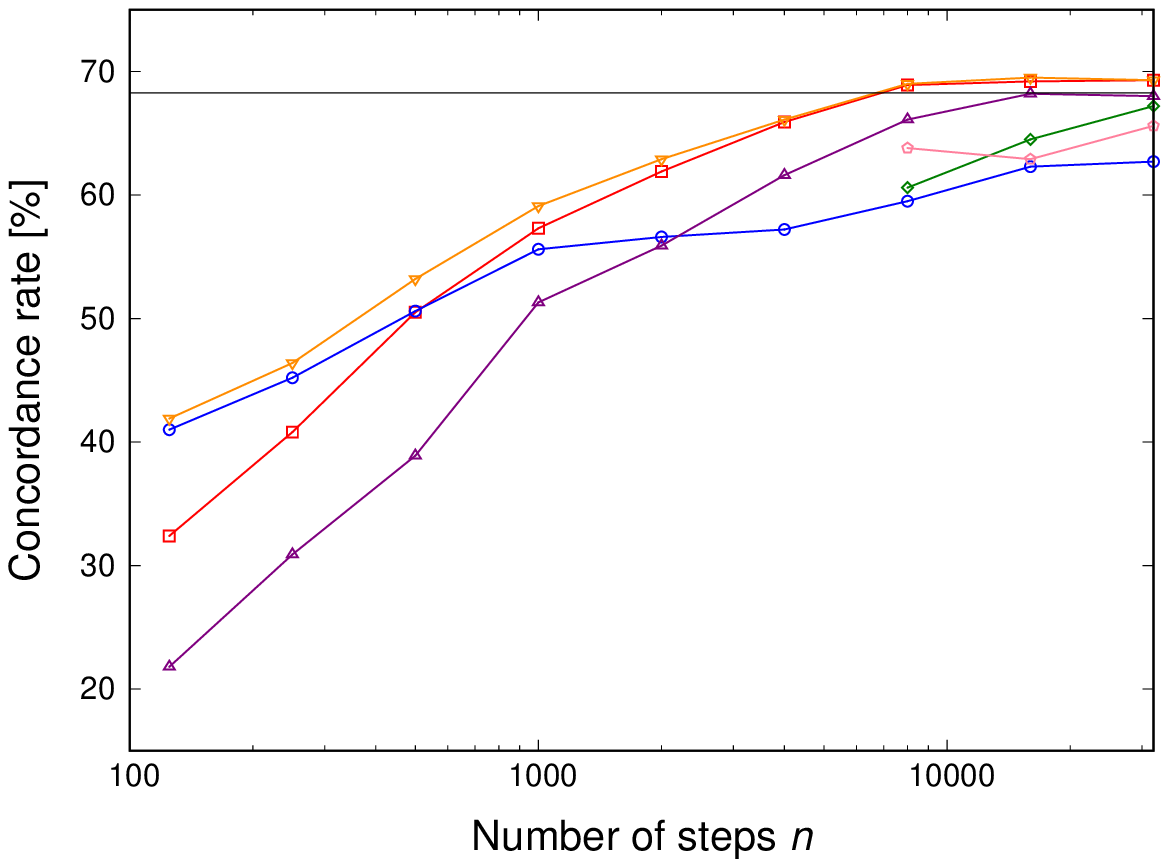}
    (c)~CCMC
  \end{minipage}
  \begin{minipage}{0.4\hsize}
    \centering
    \includegraphics[width=\hsize]{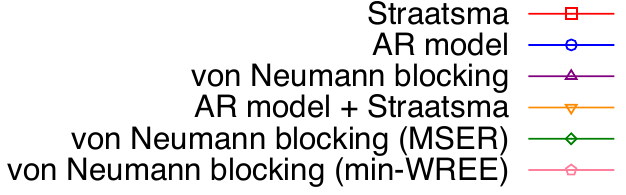}
  \end{minipage}
  \caption{
    The concordance rates between the energy means and the reference
    mean value within the errors with $1\sigma$ confidential interval 
    estimated by the error estimation methods. 
    The concordance rates are surveyed for the time-series
    generated by (a)DMC, (b)FCIQMC, and (c)CCMC,
    for different lengths of time-series.
    The black horizontal line shows 68.27~\%, which is
    the ideal value for $1\sigma$ confidence interval.
    When the measured concordance rate is closer to this value,
    the error estimation is regarded to be more reliable.
  }
  \label{fig.conc}
\end{figure*}
\begin{figure*}[htbp]
  \begin{minipage}{0.4\hsize}
    \centering
    \includegraphics[width=\hsize]{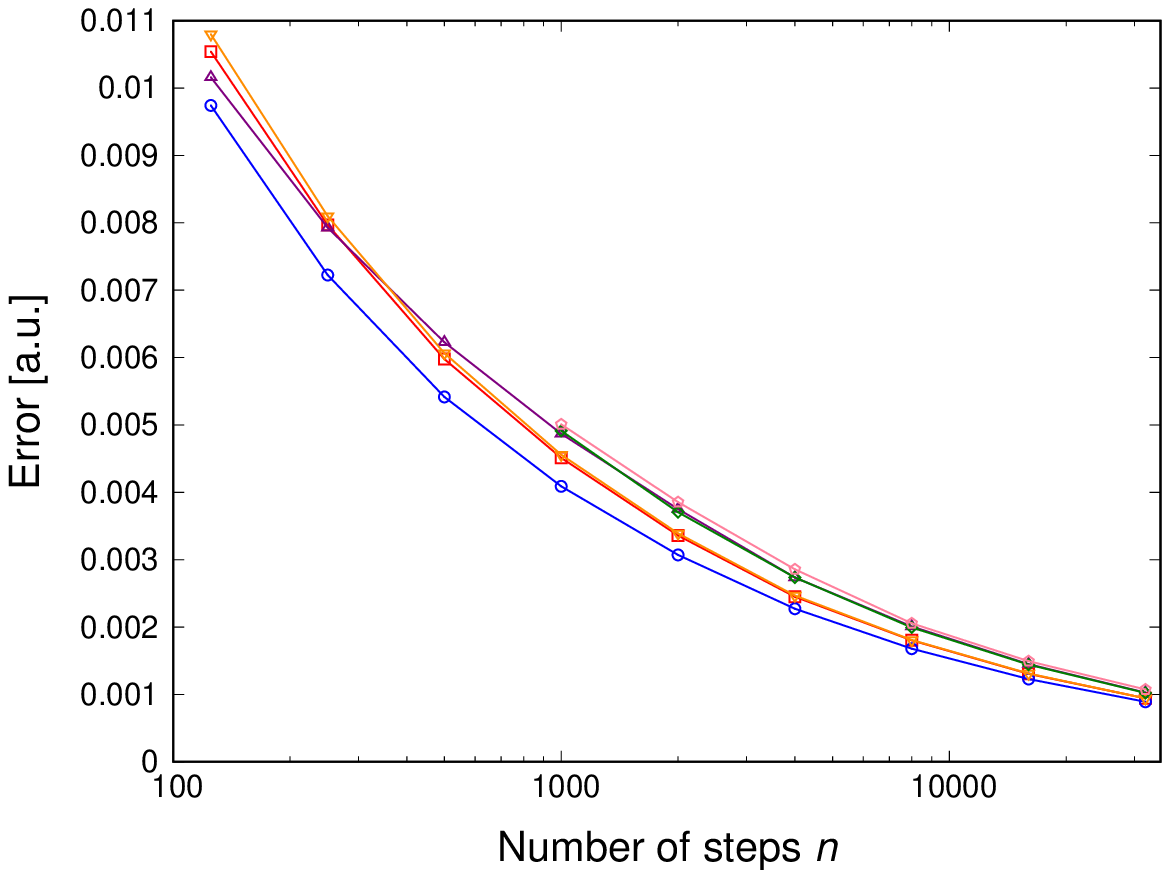}
    (a)~DMC
  \end{minipage}
  \begin{minipage}{0.4\hsize}
   \centering
    \includegraphics[width=\hsize]{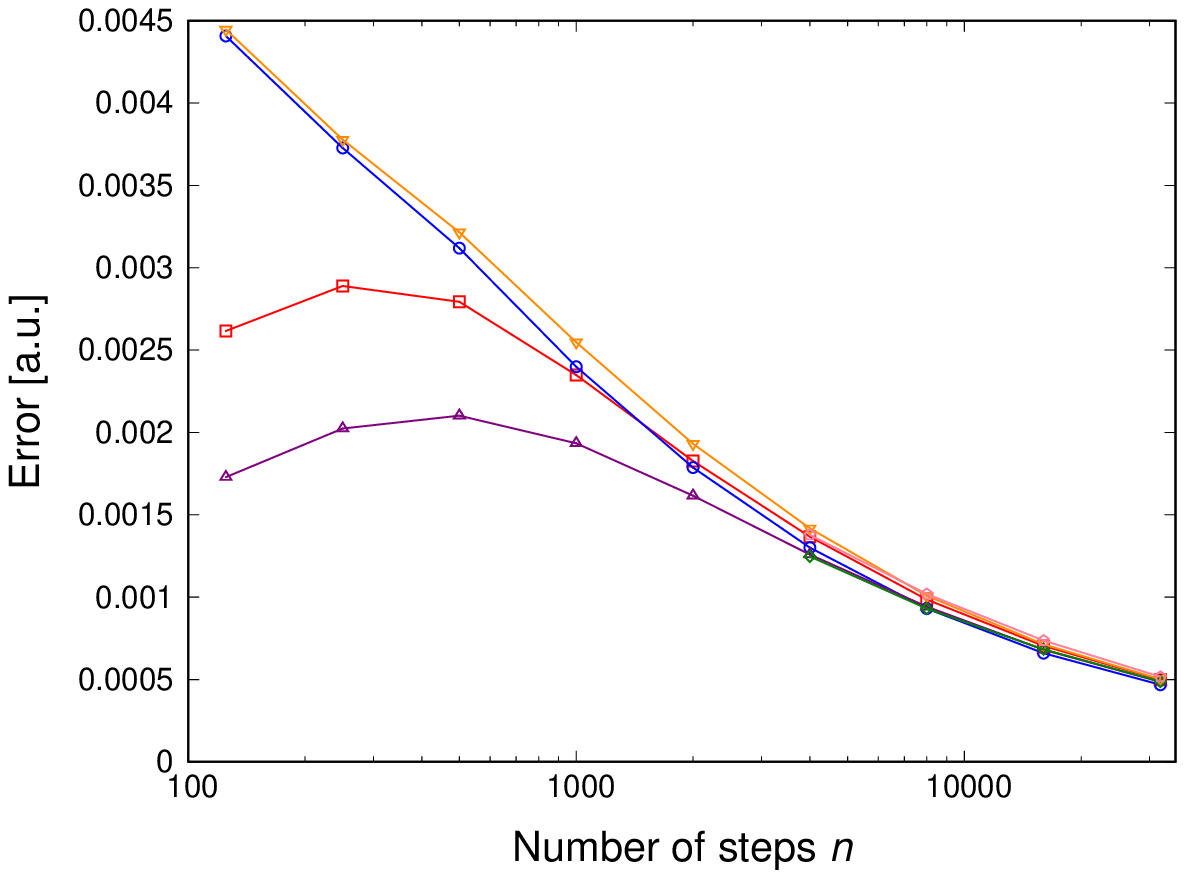}
    (b)~FCIQMC
  \end{minipage}
  \begin{minipage}{0.4\hsize}
    \centering
    \includegraphics[width=\hsize]{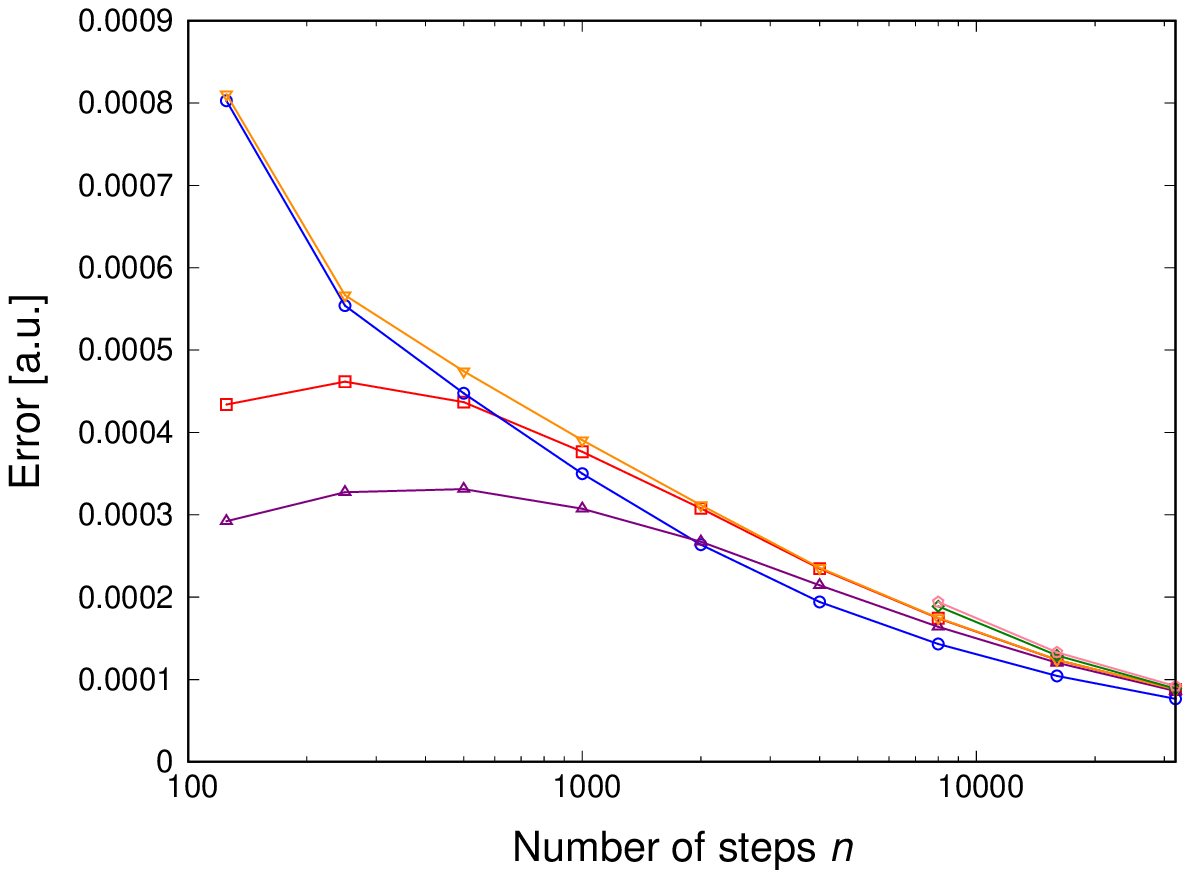}
    (c)~CCMC
  \end{minipage}
  \begin{minipage}{0.4\hsize}
    \centering
    \includegraphics[width=\hsize]{key.pdf}
  \end{minipage}
  \caption{
    The statistical errors estimated by the error estimation methods
    for time-series given by (a)DMC, (b)FCIQMC, and (c)CCMC for
    different lengths of time-series.
  }
  \label{fig.error}
\end{figure*}
\section{Estimation schemes of warm-up steps}
\label{sec.warm-up}
  We introduce two schemes for warm-up steps estimation
  in this section, MSER\cite{2008FRA} and min-WREE,
  which are implemented in HANDE code.\cite{2014SPE,2019SPE} 

\subsection*{Mean squared error rule (MSER)}
  MSER aims to give an adequate estimate
  of warm-up steps $d$ as minimizing a sum of the systematic
  error from the warm-up phase and the statistical error.
The number of warm-up steps, $d$, is determined by minimizing the following quantity:
\begin{eqnarray}
  {\rm{MSER}}\left( d \right){\rm{ = }}\frac{{s_{X}^2\left( d \right)}}{{n - d}},\;\;s_{X}^2\left( d \right) \equiv \frac{1}{{n - d}}\sum\limits_{i = 1}^{n-d} {{{\left( {{X_{i+d}} - \bar X} \right)}^2}}
  \label{eq.wsd2}
\end{eqnarray}
Here, ${s_{X}^2\left( d \right)}$ is a constant independent of $d$ corresponding to the
case  where $\left\{ {{X_i}} \right\}_{i = d}^n$ does not include a warm-up phase.
When a warm-up phase is present, ${s_{X}^2\left( d \right)}$ increases from
the constant value
according to how much warm-up phase remains in $\left\{ {{X_i}} \right\}_{i = d}^n$.
Meanwhile $1/(n-d)$ monotonically increases according to $d$, and the value of $d$ minimizing
their product gives an appropriate estimate of the number of warm-up steps.

\subsection*{Minimization of weighted relative error of the error (min-WREE)}
This scheme estimates the warm-up steps as minimizing the relative error
of error of the statistical average, ${\text{REE}}\left( d \right)$,
weighted by $1/\sqrt{n-d}$:
\begin{eqnarray}
  {{{\text{WREE}}\left( d \right) = {\text{REE}}\left( d \right)} \mathord{\left/
      {\vphantom {{{\text{WREE}}\left( d \right) = {\text{REE}}\left( d \right)} {\sqrt {n - d} }}} \right.
      \kern-\nulldelimiterspace} {\sqrt {n - d} }}
\end{eqnarray}
We evaluated ${\text{REE}}\left( d \right)$ using von Neumann blocking \cite{2011YOU} in this work.


\section{QMC calculation details}
\label{sec.details}
We employed CASINO\cite{casino_manual} for DMC calculations.
We used a Slater-Jastrow trial wave-function.\cite{2001FOU}
The determinant is generated by the Hartree--Fock method
using a STO-6G Gaussian basis set.\cite{1969HEH}
The Jastrow factor consists of one- and two-body terms
and includes 42 parameters in total.
For the DMC calculations, the target population of walkers
is set to be 1024 and the time step is 0.005~a.u.$^{-1}$.
We also used the cusp correction scheme,\cite{2005MA}
which replaces the shape of orbitals nearby ionic cores 
with Slater functions to satisfy the Kato cusp conditions.\cite{1957KAT}
  Each sample of the energy time-series is given 
  by averaging the local energies\cite{2001FOU} over
  all of the walkers for every QMC iteration.
The influence of the population fluctuation and the population
control\cite{casino_manual} is not considered in this work.

\vspace{2mm}
We performed FCIQMC and CCMC calculations with HANDE.\cite{2014SPE,2019SPE}
The reference Slater determinant is prepared by Hartree--Fock method
with cc-pVDZ Gaussian basis set\cite{1989DUN} using Psi4.\cite{2017PAR}
The target number of walker population is 500 and the time step is
2.0$\times 10^{-5}$~a.u.$^{-1}$.
  Each sample of the energy time-series is given as
  the instantaneous projected energy, which is
  a ratio between $\left\langle {{{\rm{D}}_0}\left| {\hat H} \right|{\Psi}} \right\rangle$
  and ${N_0} \equiv \left\langle {{{\rm{D}}_0}\left| {\Psi} \right.} \right\rangle$
  for every QMC iteration.
The influence of the population fluctuation and the population control
\cite{casino_manual} is not considered in this work.


\section{Results and Discussion}
\label{sec.results}
We prepared one thousand different energy time-series
for the neon atom, with the same calculation settings but
with the different random seeds, using DMC, FCIQMC, and CCMC
methods, for different lengths of time-series, respectively.
We applied the error estimation methods to them and
surveyed the concordance rate between the energy means
and the reference mean value within the estimated errors
with $1\sigma$ confidential interval (CI) as shown
in Figure~\ref{fig.conc}.
The concordance rate for a 1$\sigma$ CI is ideally 68.27\%.
Thus, when the observed rate is closer to this value,
the error estimation is regarded to be more reliable.
Here, the reference mean value is given by taking an average
of very long length of time-series, and the error is just
less than 3\% of those of the energy means of 1000 time-series.

\vspace{2mm}
First, we discuss the case of FCIQMC/CCMC (see Figure~\ref{fig.conc}bc).
All of the error estimation methods give lower concordance rate than
the ideal value for $1\sigma$ confidential interval, 68.27\%,
when the length of time-series $n$ is small.
This is typically observed for error estimation.\cite{2010THO_ar}
For comparatively short lengths of time-series,
the AR model shows the highest concordance rate among them.
The comparison of the estimated errors shown in
Figure~\ref{fig.error}bc further distinguishes the AR model
from the others: Only the AR model reproduces that
the estimated error normally decreases in proportion to $1/\sqrt{n}$.
It clearly proves the advantage of taking AR model of equation~\eqref{eq.ar1}.
On the other hand, the lowest concordance rate is measured for von Neumann blocking.
The von Neumann's criteria to check randomness and normality tends
to be not effective for small numbers of data,\cite{1978FIS}
so it underestimates the correlation length for small length of time-series.

\begin{figure}[htbp]
  \centering
  \includegraphics[width=1.0\hsize]{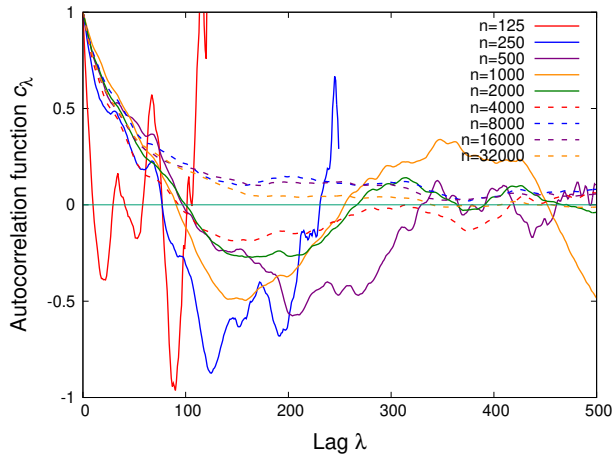}
  \caption{
    Autocorrelation functions for different length $n$
    of time-series, given by CCMC method. For comparatively
    small $n$, the autocorrelation function $c_\lambda$
    apparently includes a noise and it cannot be seen
    that $c_\lambda$ gradually converges to zero along
    with the lag $\lambda$ increasing.
  }
  \label{fig.autocorr2}
\end{figure}

\begin{figure}[htbp]
  \centering
  \includegraphics[width=1.0\hsize]{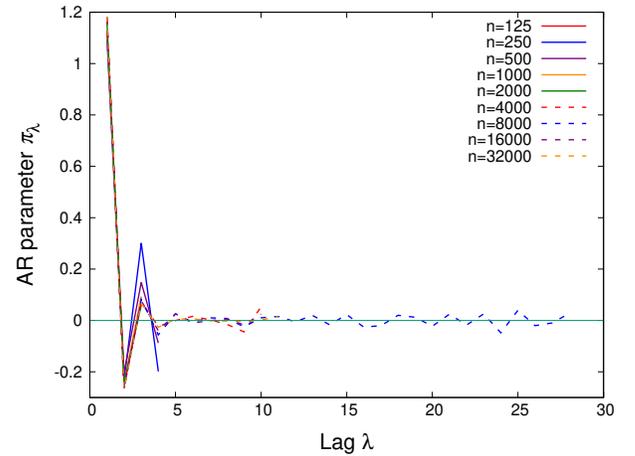}
  \caption{
    Parameters ${{\pi_\lambda}}$ in AR model fitted
    to different lengths of CCMC time-series.
    The number of the parameters is determined by AIC.\cite{1992AKA}
    The parameters are terminated or converged around zero
    within a few terms, regardless of the length of time-series.
  }
  \label{fig.pi}
\end{figure}

\vspace{2mm}
Straatsma gives the intermediate concordance rates for small lengths $n$.
The calculated $\tau$ fully depends on the autocorrelation
function $c_\lambda$ through equations~\eqref{eq.stra1} and \eqref{eq.stra3},
so we examined how the shape of the autocorrelation function $c_\lambda$ changes
according to the length $n$ of time-series in Figure~\ref{fig.autocorr2}:
The autocorrelation function $c_\lambda$ becomes negative
more quickly for smaller $n$ by the $c_\lambda$ oscillating
since it is estimated by insufficient number of terms, $n-\lambda$, through equation~\eqref{eq.stra2}.
The truncation of the sum in equation~\eqref{eq.stra3} is so drastic
that Straatsma underestimates the correlation time.
In contrast, the oscillation of $c_\lambda$ does not much affect the AR model,
although its estimation also depends on autocorrelation functions $c_\lambda$
through equation~\eqref{eq.ar2}.
This is because taking a product with ${\pi_\lambda}$ 
drastically reduces the contribution of $c_\lambda$ with large lag $\lambda$:~
Figure \ref{fig.pi} shows the expansion of the parameters ${{\pi_\lambda}}$,
where they are terminated or converged to zero only within a few terms.
Therefore, just a few terms of $c_\lambda$ from small lag $\lambda$
is used to calculate $\tau$ in AR model.


\vspace{2mm}
When the length of time-series $n$ is comparatively large,
the concordance rate of AR model converges to 68.27~\% the most slowly.
This is because the assumption of equation~\eqref{eq.ar1} in AR model cannot fully describe
the target random process, and therefore its reliability is reduced.
To summarize, the AR model (Straatsma) is the most reliable for small
(middle/large) length time-series, respectively.
We have therefore devised a hybrid scheme of AR model and Straatsma,
which works reasonably for any length of time-series.
It simply adopts the larger of the errors estimated by both methods:
$
{\sigma _{\bar X}}\left( {{\rm{hybrid}}} \right) = \max \left\{ {{\sigma _{\bar X}}\left( {{\rm{AR\,model}}} \right),{\sigma _{\bar X}}\left( {{\rm{Straatsma}}} \right)} \right\}.
$
\label{eq.max}
The concordance rate for the hybrid method is shown as `AR model + Straatsma',
in Figure \ref{fig.conc} and always comparatively close to 68.27~\%.

\begin{figure}[htbp]
  \centering
  \includegraphics[width=1.0\hsize]{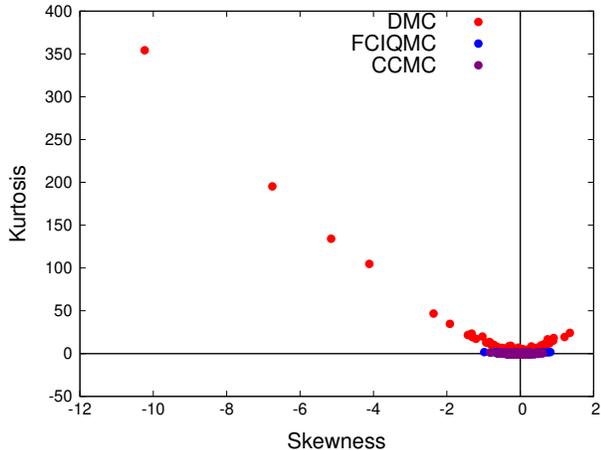}
  \caption{
    Mapping of skewness and kurtosis of the distributions 
    of 1000 time-series obtained by DMC, FCIQMC, and CCMC. 
    When the point is closer to the origin,
    the corresponding distribution is more normal.
  }
  \label{fig.skewness}
\end{figure}

\vspace{2mm}
We performed the same test of the error estimation methods
for DMC time-series. Von Neumann blocking gives the closest concordance
rate to the ideal value, 68.27\%, for any length of time-series, and
only the concordance rate of this method reaches 68.27~\%.
The difference from the case of FCIQMC/CCMC comes from that
the distribution of DMC energy time-series tends to be non-normal:
Figure \ref{fig.skewness} clearly shows that the distributions
of the time-series given by DMC have larger skewness and kurtosis
than those of FCIQMC/CCMC. As mentioned in section \ref{sec.methods},
only von Neumann blocking can take into account non-normality for error estimation,
which would be the reason why von Neumann blocking the most works in the case of DMC.
This difference in non-normality also explains
why the AR model gives the lowest concordance rate:
the AR model assumes that the randomness between the neighboring steps
is expressed by normally distributed noise, so it would not be possible to make
a description when the distribution of time-series is non-normal.

\begin{figure*}[htbp]
  \begin{minipage}{0.3\hsize}
    \centering
    \includegraphics[width=\hsize]{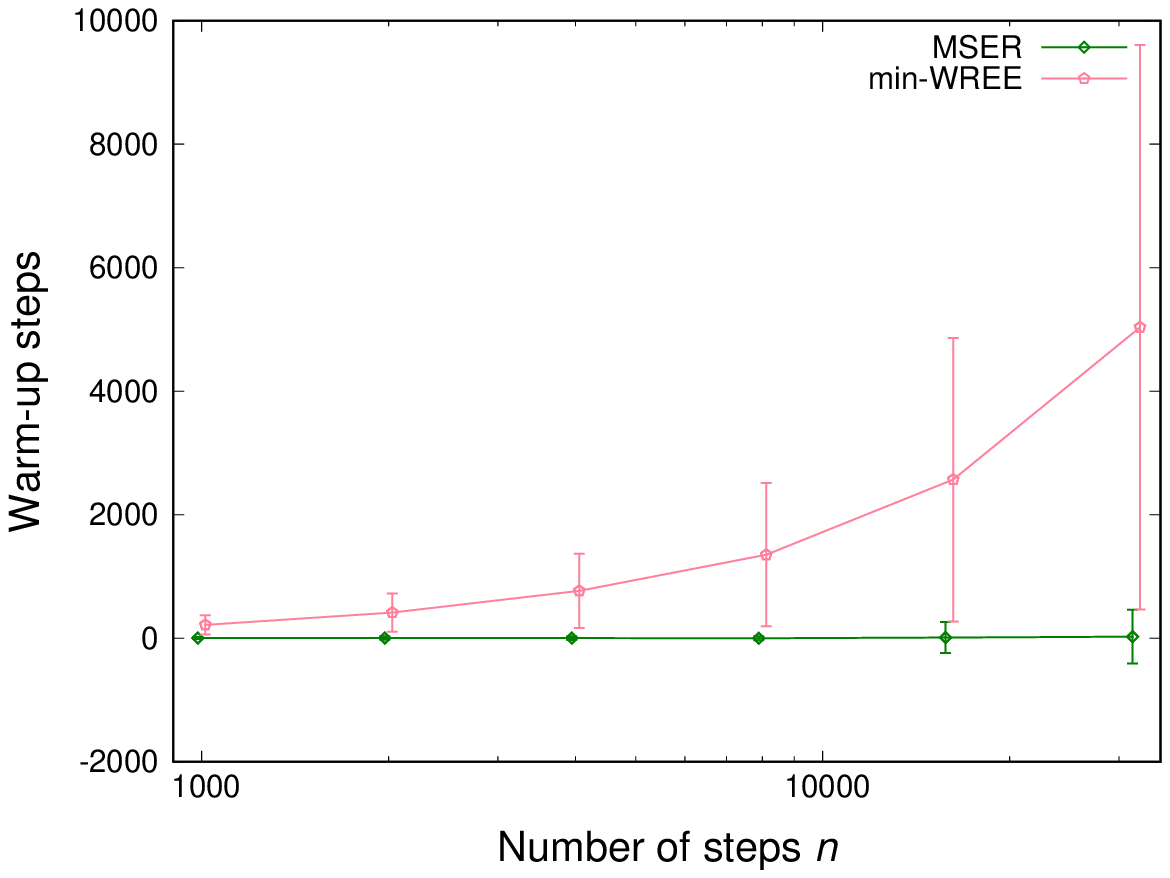}
    (a)~DMC
  \end{minipage}
  \begin{minipage}{0.3\hsize}
    \centering
    \includegraphics[width=\hsize]{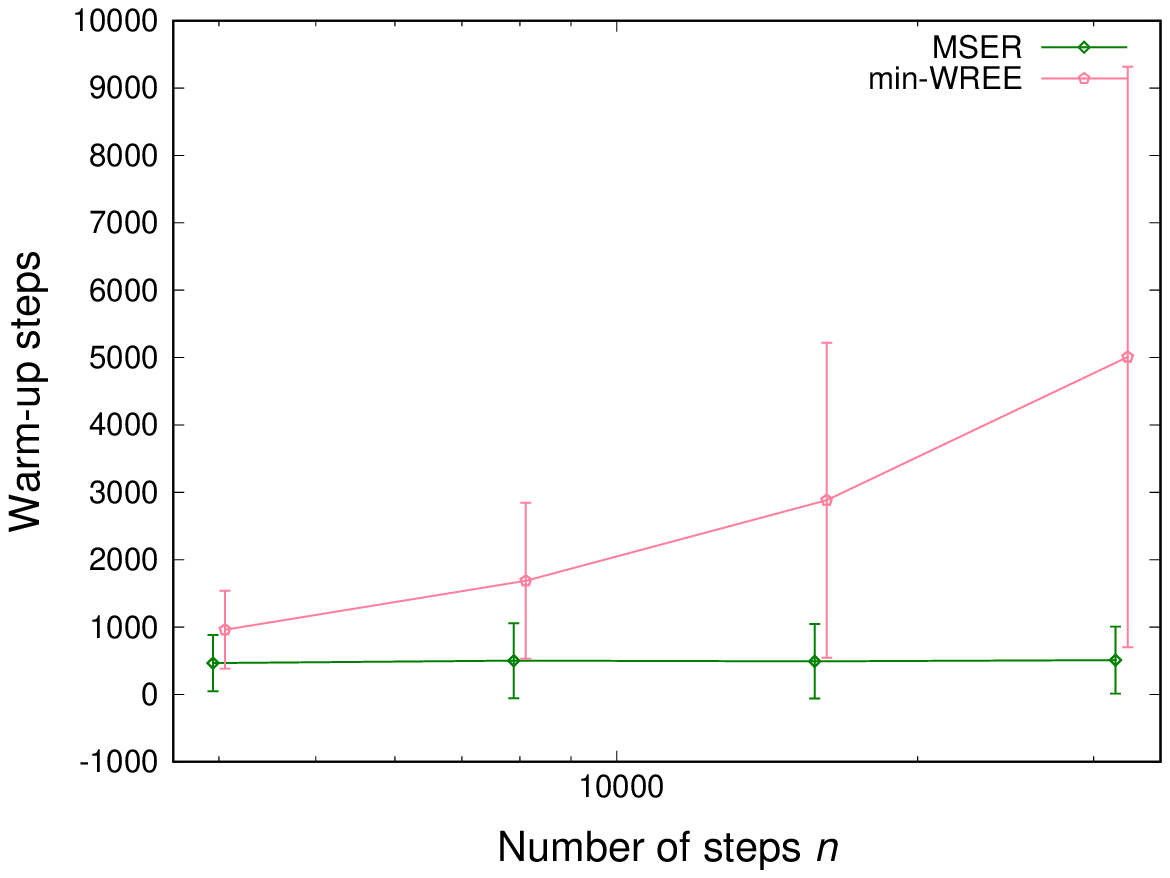}
    (b)~FCIQMC
  \end{minipage}
  \begin{minipage}{0.3\hsize}
    \centering
    \includegraphics[width=\hsize]{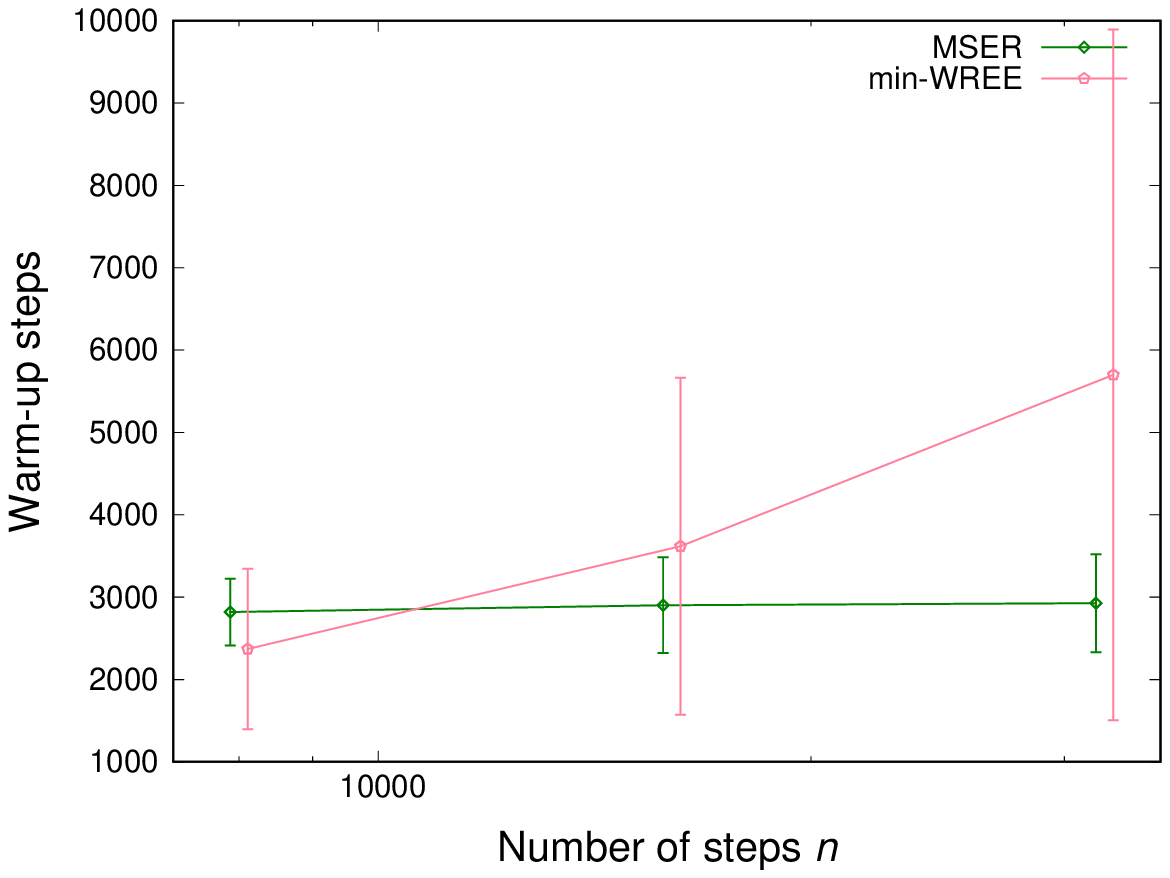}
    (c)~CCMC
  \end{minipage}
  \caption{  \label{fig.init}
    The warm-up steps estimated by MSER and
    min-WREE for the time-series generated by
    DMC, FCIQMC, and CCMC. The error bars correspond
    to the statistical errors.
  }
\end{figure*}

\vspace{2mm}
Finally, we discuss the performance of the estimation schemes
of warm-up steps, MSER and min-WREE.
We apply these schemes to the time-series including non-plateau part
and removed the estimated warm-up steps $d$.
Then, we applied `von Neumann blocking' to obtain the concordance rate
and the statistical error.
Figure~\ref{fig.conc} shows that MSER basically
gives higher concordance rate and smaller statistical error
especially in the case of DMC: MSER is superior
to min-WREE on the whole. The advantage is further
distinguished comparing the average of warm-up steps.
They are shown in Figure~\ref{fig.init} with the standard errors.
It clearly shows that the estimations of min-WREE 
strongly depends on the length of time-series and largely
scattered. On the other hand, MSER estimates
constant warm-up steps with small variances, independent of
the length of time-series.
%

\section{Conclusion}
\label{sec.conclusion}
We compared the reliability of three kinds of error estimation methods,
Straatsma,\cite{1986STR} AR model,\cite{2010THO_ar} and von Neumann blocking,\cite{1978FIS,2011YOU}
in a statistical manner, when they are applied to the energy time-series given
by applying DMC\cite{2001FOU}, FCIQMC\cite{2010BOO}, and CCMC\cite{2010THO_ccmc,2016SPE}
to the neon atom.
In the case of FCIQMC/CCMC, it is shown that Straatsma (the AR model) is the most reliable
for comparatively long (short) length of time-series, respectively.
We established that the assumption in the AR model significantly reduces 
the influence of the oscillation of autocorrelation functions for short
time-series lengths, but we concluded that the systematic error from the assumption
is pronounced for long lengths.
We devised a hybrid scheme, which takes the larger error from
the ones estimated by Straatsma and AR model, and established it works for any length of time-series.
In contrast, for DMC, we showed that von Neumann blocking is
the most reliable for any length of time-series.
This method has an advantage of considering the non-normality of the distribution
of time-series, and we showed an strong evidence that the advantage is essential
to analyse the DMC results.
We have also tested two kinds of warm-up steps estimation schemes,
MSER\cite{2008FRA} and min-WREE\cite{2014SPE},
and established that MSER gives constant and reasonable
estimations of warm-up steps, independent of the length and for
time-series made by any of the QMC methods.

\section*{Acknowledgments}
The computation in this work has been performed using the facilities
of the Research Center for Advanced Computing Infrastructure (RCACI) at JAIST.
T.I. is grateful for financial suport from Grant-in-Aid for JSPS Research Fellow (18J12653).
K.H. is grateful for financial support from a KAKENHI grant (JP17K17762),
a Grant-in-Aid for Scientific Research on Innovative Areas ``Mixed Anion'' project (JP16H06439) from MEXT, 
PRESTO (JPMJPR16NA) and the Materials research by Information Integration Initiative (MI$^2$I) project 
of the Support Program for Starting Up Innovation Hub from Japan Science and Technology Agency (JST). 
R.M. is grateful for financial supports from MEXT-KAKENHI (17H05478 and 16KK0097), 
from Toyota Motor Corporation, from I-O DATA Foundation, 
and from the Air Force Office of Scientific Research (AFOSR-AOARD/FA2386-17-1-4049).
R.M. and K.H. are also grateful to financial supports from MEXT-FLAGSHIP2020 (hp170269, hp170220).
A.J.W.T. thanks the Royal Society for a University Research
Fellowship (UF110161).
\bibliography{references}

\begin{thebibliography}{36}%
\makeatletter
\providecommand \@ifxundefined [1]{%
 \@ifx{#1\undefined}
}%
\providecommand \@ifnum [1]{%
 \ifnum #1\expandafter \@firstoftwo
 \else \expandafter \@secondoftwo
 \fi
}%
\providecommand \@ifx [1]{%
 \ifx #1\expandafter \@firstoftwo
 \else \expandafter \@secondoftwo
 \fi
}%
\providecommand \natexlab [1]{#1}%
\providecommand \enquote  [1]{``#1''}%
\providecommand \bibnamefont  [1]{#1}%
\providecommand \bibfnamefont [1]{#1}%
\providecommand \citenamefont [1]{#1}%
\providecommand \href@noop [0]{\@secondoftwo}%
\providecommand \href [0]{\begingroup \@sanitize@url \@href}%
\providecommand \@href[1]{\@@startlink{#1}\@@href}%
\providecommand \@@href[1]{\endgroup#1\@@endlink}%
\providecommand \@sanitize@url [0]{\catcode `\\12\catcode `\$12\catcode
  `\&12\catcode `\#12\catcode `\^12\catcode `\_12\catcode `\%12\relax}%
\providecommand \@@startlink[1]{}%
\providecommand \@@endlink[0]{}%
\providecommand \url  [0]{\begingroup\@sanitize@url \@url }%
\providecommand \@url [1]{\endgroup\@href {#1}{\urlprefix }}%
\providecommand \urlprefix  [0]{URL }%
\providecommand \Eprint [0]{\href }%
\providecommand \doibase [0]{http://dx.doi.org/}%
\providecommand \selectlanguage [0]{\@gobble}%
\providecommand \bibinfo  [0]{\@secondoftwo}%
\providecommand \bibfield  [0]{\@secondoftwo}%
\providecommand \translation [1]{[#1]}%
\providecommand \BibitemOpen [0]{}%
\providecommand \bibitemStop [0]{}%
\providecommand \bibitemNoStop [0]{.\EOS\space}%
\providecommand \EOS [0]{\spacefactor3000\relax}%
\providecommand \BibitemShut  [1]{\csname bibitem#1\endcsname}%
\let\auto@bib@innerbib\@empty
\bibitem [{\citenamefont {Spencer}\ \emph {et~al.}(2018)\citenamefont
  {Spencer}, \citenamefont {Neufeld}, \citenamefont {Vigor}, \citenamefont
  {Franklin},\ and\ \citenamefont {Thom}}]{2018SPE}%
  \BibitemOpen
  \bibfield  {author} {\bibinfo {author} {\bibfnamefont {J.~S.}\ \bibnamefont
  {Spencer}}, \bibinfo {author} {\bibfnamefont {V.~A.}\ \bibnamefont
  {Neufeld}}, \bibinfo {author} {\bibfnamefont {W.~A.}\ \bibnamefont {Vigor}},
  \bibinfo {author} {\bibfnamefont {R.~S.~T.}\ \bibnamefont {Franklin}}, \ and\
  \bibinfo {author} {\bibfnamefont {A.~J.~W.}\ \bibnamefont {Thom}},\ }\href
  {\doibase 10.1063/1.5047420} {\bibfield  {journal} {\bibinfo  {journal} {The
  Journal of Chemical Physics}\ }\textbf {\bibinfo {volume} {149}},\ \bibinfo
  {pages} {204103} (\bibinfo {year} {2018})},\ \Eprint
  {http://arxiv.org/abs/https://doi.org/10.1063/1.5047420}
  {https://doi.org/10.1063/1.5047420} \BibitemShut {NoStop}%
\bibitem [{\citenamefont {Blunt}\ \emph {et~al.}(2015)\citenamefont {Blunt},
  \citenamefont {Smart}, \citenamefont {Kersten}, \citenamefont {Spencer},
  \citenamefont {Booth},\ and\ \citenamefont {Alavi}}]{2015BLU}%
  \BibitemOpen
  \bibfield  {author} {\bibinfo {author} {\bibfnamefont {N.~S.}\ \bibnamefont
  {Blunt}}, \bibinfo {author} {\bibfnamefont {S.~D.}\ \bibnamefont {Smart}},
  \bibinfo {author} {\bibfnamefont {J.~A.~F.}\ \bibnamefont {Kersten}},
  \bibinfo {author} {\bibfnamefont {J.~S.}\ \bibnamefont {Spencer}}, \bibinfo
  {author} {\bibfnamefont {G.~H.}\ \bibnamefont {Booth}}, \ and\ \bibinfo
  {author} {\bibfnamefont {A.}~\bibnamefont {Alavi}},\ }\href {\doibase
  10.1063/1.4920975} {\bibfield  {journal} {\bibinfo  {journal} {The Journal of
  Chemical Physics}\ }\textbf {\bibinfo {volume} {142}},\ \bibinfo {pages}
  {184107} (\bibinfo {year} {2015})},\ \Eprint
  {http://arxiv.org/abs/https://doi.org/10.1063/1.4920975}
  {https://doi.org/10.1063/1.4920975} \BibitemShut {NoStop}%
\bibitem [{\citenamefont {Mathuriya}\ \emph {et~al.}(2017)\citenamefont
  {Mathuriya}, \citenamefont {Luo}, \citenamefont {Clay}, \citenamefont
  {Benali}, \citenamefont {Shulenburger},\ and\ \citenamefont {Kim}}]{2017MAT}%
  \BibitemOpen
  \bibfield  {author} {\bibinfo {author} {\bibfnamefont {A.}~\bibnamefont
  {Mathuriya}}, \bibinfo {author} {\bibfnamefont {Y.}~\bibnamefont {Luo}},
  \bibinfo {author} {\bibfnamefont {R.~C.}\ \bibnamefont {Clay}, \bibfnamefont
  {III}}, \bibinfo {author} {\bibfnamefont {A.}~\bibnamefont {Benali}},
  \bibinfo {author} {\bibfnamefont {L.}~\bibnamefont {Shulenburger}}, \ and\
  \bibinfo {author} {\bibfnamefont {J.}~\bibnamefont {Kim}},\ }in\ \href
  {\doibase 10.1145/3126908.3126952} {\emph {\bibinfo {booktitle} {Proceedings
  of the International Conference for High Performance Computing, Networking,
  Storage and Analysis}}},\ \bibinfo {series and number} {SC '17}\ (\bibinfo
  {publisher} {ACM},\ \bibinfo {address} {New York, NY, USA},\ \bibinfo {year}
  {2017})\ pp.\ \bibinfo {pages} {38:1--38:12}\BibitemShut {NoStop}%
\bibitem [{\citenamefont {Ichibha}\ \emph {et~al.}(2017)\citenamefont
  {Ichibha}, \citenamefont {Hou}, \citenamefont {Hongo},\ and\ \citenamefont
  {Maezono}}]{2017ICH}%
  \BibitemOpen
  \bibfield  {author} {\bibinfo {author} {\bibfnamefont {T.}~\bibnamefont
  {Ichibha}}, \bibinfo {author} {\bibfnamefont {Z.}~\bibnamefont {Hou}},
  \bibinfo {author} {\bibfnamefont {K.}~\bibnamefont {Hongo}}, \ and\ \bibinfo
  {author} {\bibfnamefont {R.}~\bibnamefont {Maezono}},\ }\href {\doibase
  10.1038/s41598-017-01668-6} {\bibfield  {journal} {\bibinfo  {journal} {Sci.
  Rep.}\ }\textbf {\bibinfo {volume} {7}},\ \bibinfo {pages} {2011} (\bibinfo
  {year} {2017})}\BibitemShut {NoStop}%
\bibitem [{\citenamefont {Luo}\ \emph {et~al.}(2016)\citenamefont {Luo},
  \citenamefont {Benali}, \citenamefont {Shulenburger}, \citenamefont {Krogel},
  \citenamefont {Heinonen},\ and\ \citenamefont {Kent}}]{2016LUO}%
  \BibitemOpen
  \bibfield  {author} {\bibinfo {author} {\bibfnamefont {Y.}~\bibnamefont
  {Luo}}, \bibinfo {author} {\bibfnamefont {A.}~\bibnamefont {Benali}},
  \bibinfo {author} {\bibfnamefont {L.}~\bibnamefont {Shulenburger}}, \bibinfo
  {author} {\bibfnamefont {J.~T.}\ \bibnamefont {Krogel}}, \bibinfo {author}
  {\bibfnamefont {O.}~\bibnamefont {Heinonen}}, \ and\ \bibinfo {author}
  {\bibfnamefont {P.~R.~C.}\ \bibnamefont {Kent}},\ }\href
  {http://stacks.iop.org/1367-2630/18/i=11/a=113049} {\bibfield  {journal}
  {\bibinfo  {journal} {New Journal of Physics}\ }\textbf {\bibinfo {volume}
  {18}},\ \bibinfo {pages} {113049} (\bibinfo {year} {2016})}\BibitemShut
  {NoStop}%
\bibitem [{\citenamefont {Trail}\ \emph {et~al.}(2017)\citenamefont {Trail},
  \citenamefont {Monserrat}, \citenamefont {L\'opez~R\'{\i}os}, \citenamefont
  {Maezono},\ and\ \citenamefont {Needs}}]{2017TRA}%
  \BibitemOpen
  \bibfield  {author} {\bibinfo {author} {\bibfnamefont {J.}~\bibnamefont
  {Trail}}, \bibinfo {author} {\bibfnamefont {B.}~\bibnamefont {Monserrat}},
  \bibinfo {author} {\bibfnamefont {P.}~\bibnamefont {L\'opez~R\'{\i}os}},
  \bibinfo {author} {\bibfnamefont {R.}~\bibnamefont {Maezono}}, \ and\
  \bibinfo {author} {\bibfnamefont {R.~J.}\ \bibnamefont {Needs}},\ }\href
  {\doibase 10.1103/PhysRevB.95.121108} {\bibfield  {journal} {\bibinfo
  {journal} {Phys. Rev. B}\ }\textbf {\bibinfo {volume} {95}},\ \bibinfo
  {pages} {121108} (\bibinfo {year} {2017})}\BibitemShut {NoStop}%
\bibitem [{\citenamefont {Ken}\ \emph {et~al.}(2018)\citenamefont {Ken},
  \citenamefont {Ichibha}, \citenamefont {Hongo},\ and\ \citenamefont
  {Maezono}}]{2018KEN}%
  \BibitemOpen
  \bibfield  {author} {\bibinfo {author} {\bibfnamefont {Q.~S.}\ \bibnamefont
  {Ken}}, \bibinfo {author} {\bibfnamefont {T.}~\bibnamefont {Ichibha}},
  \bibinfo {author} {\bibfnamefont {K.}~\bibnamefont {Hongo}}, \ and\ \bibinfo
  {author} {\bibfnamefont {R.}~\bibnamefont {Maezono}},\ }\href@noop {}
  {\enquote {\bibinfo {title} {Difficulty to capture non-additive enhancement
  of stacking energy by conventional ab initio methods},}\ } (\bibinfo {year}
  {2018}),\ \Eprint {http://arxiv.org/abs/arXiv:1807.04168} {arXiv:1807.04168}
  \BibitemShut {NoStop}%
\bibitem [{\citenamefont {Benali}\ \emph {et~al.}(2018)\citenamefont {Benali},
  \citenamefont {Luo}, \citenamefont {Shin}, \citenamefont {Pahls},\ and\
  \citenamefont {Heinonen}}]{2018BEN}%
  \BibitemOpen
  \bibfield  {author} {\bibinfo {author} {\bibfnamefont {A.}~\bibnamefont
  {Benali}}, \bibinfo {author} {\bibfnamefont {Y.}~\bibnamefont {Luo}},
  \bibinfo {author} {\bibfnamefont {H.}~\bibnamefont {Shin}}, \bibinfo {author}
  {\bibfnamefont {D.}~\bibnamefont {Pahls}}, \ and\ \bibinfo {author}
  {\bibfnamefont {O.}~\bibnamefont {Heinonen}},\ }\href {\doibase
  10.1021/acs.jpcc.8b02368} {\bibfield  {journal} {\bibinfo  {journal} {The
  Journal of Physical Chemistry C}\ }\textbf {\bibinfo {volume} {122}},\
  \bibinfo {pages} {16683} (\bibinfo {year} {2018})},\ \Eprint
  {http://arxiv.org/abs/https://doi.org/10.1021/acs.jpcc.8b02368}
  {https://doi.org/10.1021/acs.jpcc.8b02368} \BibitemShut {NoStop}%
\bibitem [{\citenamefont {Foulkes}\ \emph {et~al.}(2001)\citenamefont
  {Foulkes}, \citenamefont {Mitas}, \citenamefont {Needs},\ and\ \citenamefont
  {Rajagopal}}]{2001FOU}%
  \BibitemOpen
  \bibfield  {author} {\bibinfo {author} {\bibfnamefont {W.~M.~C.}\
  \bibnamefont {Foulkes}}, \bibinfo {author} {\bibfnamefont {L.}~\bibnamefont
  {Mitas}}, \bibinfo {author} {\bibfnamefont {R.~J.}\ \bibnamefont {Needs}}, \
  and\ \bibinfo {author} {\bibfnamefont {G.}~\bibnamefont {Rajagopal}},\ }\href
  {\doibase 10.1103/RevModPhys.73.33} {\bibfield  {journal} {\bibinfo
  {journal} {Rev. Mod. Phys.}\ }\textbf {\bibinfo {volume} {73}},\ \bibinfo
  {pages} {33} (\bibinfo {year} {2001})}\BibitemShut {NoStop}%
\bibitem [{\citenamefont {Canc\`es}\ \emph {et~al.}(2006)\citenamefont
  {Canc\`es}, \citenamefont {Jourdain},\ and\ \citenamefont
  {Leli\`evre}}]{2006CAN}%
  \BibitemOpen
  \bibfield  {author} {\bibinfo {author} {\bibfnamefont {E.}~\bibnamefont
  {Canc\`es}}, \bibinfo {author} {\bibfnamefont {B.}~\bibnamefont {Jourdain}},
  \ and\ \bibinfo {author} {\bibfnamefont {T.}~\bibnamefont {Leli\`evre}},\
  }\href {\doibase 10.1142/S0218202506001583} {\bibfield  {journal} {\bibinfo
  {journal} {Mathematical Models and Methods in Applied Sciences}\ }\textbf
  {\bibinfo {volume} {16}},\ \bibinfo {pages} {1403} (\bibinfo {year}
  {2006})},\ \Eprint
  {http://arxiv.org/abs/https://doi.org/10.1142/S0218202506001583}
  {https://doi.org/10.1142/S0218202506001583} \BibitemShut {NoStop}%
\bibitem [{\citenamefont {L\'opez~R\'{\i}os}\ \emph {et~al.}(2006)\citenamefont
  {L\'opez~R\'{\i}os}, \citenamefont {Ma}, \citenamefont {Drummond},
  \citenamefont {Towler},\ and\ \citenamefont {Needs}}]{2006RIO}%
  \BibitemOpen
  \bibfield  {author} {\bibinfo {author} {\bibfnamefont {P.}~\bibnamefont
  {L\'opez~R\'{\i}os}}, \bibinfo {author} {\bibfnamefont {A.}~\bibnamefont
  {Ma}}, \bibinfo {author} {\bibfnamefont {N.~D.}\ \bibnamefont {Drummond}},
  \bibinfo {author} {\bibfnamefont {M.~D.}\ \bibnamefont {Towler}}, \ and\
  \bibinfo {author} {\bibfnamefont {R.~J.}\ \bibnamefont {Needs}},\ }\href
  {\doibase 10.1103/PhysRevE.74.066701} {\bibfield  {journal} {\bibinfo
  {journal} {Phys. Rev. E}\ }\textbf {\bibinfo {volume} {74}},\ \bibinfo
  {pages} {066701} (\bibinfo {year} {2006})}\BibitemShut {NoStop}%
\bibitem [{\citenamefont {Morales}\ \emph {et~al.}(2012)\citenamefont
  {Morales}, \citenamefont {McMinis}, \citenamefont {Clark}, \citenamefont
  {Kim},\ and\ \citenamefont {Scuseria}}]{2012MOR}%
  \BibitemOpen
  \bibfield  {author} {\bibinfo {author} {\bibfnamefont {M.~A.}\ \bibnamefont
  {Morales}}, \bibinfo {author} {\bibfnamefont {J.}~\bibnamefont {McMinis}},
  \bibinfo {author} {\bibfnamefont {B.~K.}\ \bibnamefont {Clark}}, \bibinfo
  {author} {\bibfnamefont {J.}~\bibnamefont {Kim}}, \ and\ \bibinfo {author}
  {\bibfnamefont {G.~E.}\ \bibnamefont {Scuseria}},\ }\href {\doibase
  10.1021/ct3003404} {\bibfield  {journal} {\bibinfo  {journal} {Journal of
  Chemical Theory and Computation}\ }\textbf {\bibinfo {volume} {8}},\ \bibinfo
  {pages} {2181} (\bibinfo {year} {2012})},\ \bibinfo {note} {pMID: 26588949},\
  \Eprint {http://arxiv.org/abs/https://doi.org/10.1021/ct3003404}
  {https://doi.org/10.1021/ct3003404} \BibitemShut {NoStop}%
\bibitem [{\citenamefont {Booth}\ and\ \citenamefont {Alavi}(2010)}]{2010BOO}%
  \BibitemOpen
  \bibfield  {author} {\bibinfo {author} {\bibfnamefont {G.~H.}\ \bibnamefont
  {Booth}}\ and\ \bibinfo {author} {\bibfnamefont {A.}~\bibnamefont {Alavi}},\
  }\href {\doibase 10.1063/1.3407895} {\bibfield  {journal} {\bibinfo
  {journal} {The Journal of Chemical Physics}\ }\textbf {\bibinfo {volume}
  {132}},\ \bibinfo {pages} {174104} (\bibinfo {year} {2010})},\ \Eprint
  {http://arxiv.org/abs/https://doi.org/10.1063/1.3407895}
  {https://doi.org/10.1063/1.3407895} \BibitemShut {NoStop}%
\bibitem [{\citenamefont {Cleland}\ \emph {et~al.}(2010)\citenamefont
  {Cleland}, \citenamefont {Booth},\ and\ \citenamefont {Alavi}}]{2010CLE}%
  \BibitemOpen
  \bibfield  {author} {\bibinfo {author} {\bibfnamefont {D.}~\bibnamefont
  {Cleland}}, \bibinfo {author} {\bibfnamefont {G.~H.}\ \bibnamefont {Booth}},
  \ and\ \bibinfo {author} {\bibfnamefont {A.}~\bibnamefont {Alavi}},\ }\href
  {\doibase 10.1063/1.3302277} {\bibfield  {journal} {\bibinfo  {journal} {The
  Journal of Chemical Physics}\ }\textbf {\bibinfo {volume} {132}},\ \bibinfo
  {pages} {041103} (\bibinfo {year} {2010})},\ \Eprint
  {http://arxiv.org/abs/https://doi.org/10.1063/1.3302277}
  {https://doi.org/10.1063/1.3302277} \BibitemShut {NoStop}%
\bibitem [{\citenamefont {Cleland}\ \emph {et~al.}(2011)\citenamefont
  {Cleland}, \citenamefont {Booth},\ and\ \citenamefont {Alavi}}]{2011CLE}%
  \BibitemOpen
  \bibfield  {author} {\bibinfo {author} {\bibfnamefont {D.~M.}\ \bibnamefont
  {Cleland}}, \bibinfo {author} {\bibfnamefont {G.~H.}\ \bibnamefont {Booth}},
  \ and\ \bibinfo {author} {\bibfnamefont {A.}~\bibnamefont {Alavi}},\ }\href
  {\doibase 10.1063/1.3525712} {\bibfield  {journal} {\bibinfo  {journal} {The
  Journal of Chemical Physics}\ }\textbf {\bibinfo {volume} {134}},\ \bibinfo
  {pages} {024112} (\bibinfo {year} {2011})},\ \Eprint
  {http://arxiv.org/abs/https://doi.org/10.1063/1.3525712}
  {https://doi.org/10.1063/1.3525712} \BibitemShut {NoStop}%
\bibitem [{\citenamefont {Scott}\ and\ \citenamefont {Thom}(2017)}]{2017SCO}%
  \BibitemOpen
  \bibfield  {author} {\bibinfo {author} {\bibfnamefont {C.~J.~C.}\
  \bibnamefont {Scott}}\ and\ \bibinfo {author} {\bibfnamefont {A.~J.~W.}\
  \bibnamefont {Thom}},\ }\href {\doibase 10.1063/1.4991795} {\bibfield
  {journal} {\bibinfo  {journal} {The Journal of Chemical Physics}\ }\textbf
  {\bibinfo {volume} {147}},\ \bibinfo {pages} {124105} (\bibinfo {year}
  {2017})},\ \Eprint {http://arxiv.org/abs/https://doi.org/10.1063/1.4991795}
  {https://doi.org/10.1063/1.4991795} \BibitemShut {NoStop}%
\bibitem [{\citenamefont {Thom}(2010)}]{2010THO_ccmc}%
  \BibitemOpen
  \bibfield  {author} {\bibinfo {author} {\bibfnamefont {A.~J.~W.}\
  \bibnamefont {Thom}},\ }\href {\doibase 10.1103/PhysRevLett.105.263004}
  {\bibfield  {journal} {\bibinfo  {journal} {Phys. Rev. Lett.}\ }\textbf
  {\bibinfo {volume} {105}},\ \bibinfo {pages} {263004} (\bibinfo {year}
  {2010})}\BibitemShut {NoStop}%
\bibitem [{\citenamefont {Spencer}\ and\ \citenamefont {Thom}(2016)}]{2016SPE}%
  \BibitemOpen
  \bibfield  {author} {\bibinfo {author} {\bibfnamefont {J.~S.}\ \bibnamefont
  {Spencer}}\ and\ \bibinfo {author} {\bibfnamefont {A.~J.~W.}\ \bibnamefont
  {Thom}},\ }\href {\doibase 10.1063/1.4942173} {\bibfield  {journal} {\bibinfo
   {journal} {The Journal of Chemical Physics}\ }\textbf {\bibinfo {volume}
  {144}},\ \bibinfo {pages} {084108} (\bibinfo {year} {2016})},\ \Eprint
  {http://arxiv.org/abs/https://doi.org/10.1063/1.4942173}
  {https://doi.org/10.1063/1.4942173} \BibitemShut {NoStop}%
\bibitem [{\citenamefont {Alexopoulos}\ \emph {et~al.}(2016)\citenamefont
  {Alexopoulos}, \citenamefont {Goldsman}, \citenamefont {Tang},\ and\
  \citenamefont {Wilson}}]{2017ALE}%
  \BibitemOpen
  \bibfield  {author} {\bibinfo {author} {\bibfnamefont {C.}~\bibnamefont
  {Alexopoulos}}, \bibinfo {author} {\bibfnamefont {D.}~\bibnamefont
  {Goldsman}}, \bibinfo {author} {\bibfnamefont {P.}~\bibnamefont {Tang}}, \
  and\ \bibinfo {author} {\bibfnamefont {J.~R.}\ \bibnamefont {Wilson}},\
  }\href {\doibase 10.1080/0740817X.2016.1163443} {\bibfield  {journal}
  {\bibinfo  {journal} {IIE Transactions}\ }\textbf {\bibinfo {volume} {48}},\
  \bibinfo {pages} {864} (\bibinfo {year} {2016})},\ \Eprint
  {http://arxiv.org/abs/https://doi.org/10.1080/0740817X.2016.1163443}
  {https://doi.org/10.1080/0740817X.2016.1163443} \BibitemShut {NoStop}%
\bibitem [{\citenamefont {Straatsma}\ \emph {et~al.}(1986)\citenamefont
  {Straatsma}, \citenamefont {Berendsen},\ and\ \citenamefont
  {Stam}}]{1986STR}%
  \BibitemOpen
  \bibfield  {author} {\bibinfo {author} {\bibfnamefont {T.}~\bibnamefont
  {Straatsma}}, \bibinfo {author} {\bibfnamefont {H.}~\bibnamefont
  {Berendsen}}, \ and\ \bibinfo {author} {\bibfnamefont {A.}~\bibnamefont
  {Stam}},\ }\href {\doibase 10.1080/00268978600100071} {\bibfield  {journal}
  {\bibinfo  {journal} {Molecular Physics}\ }\textbf {\bibinfo {volume} {57}},\
  \bibinfo {pages} {89} (\bibinfo {year} {1986})},\ \Eprint
  {http://arxiv.org/abs/https://doi.org/10.1080/00268978600100071}
  {https://doi.org/10.1080/00268978600100071} \BibitemShut {NoStop}%
\bibitem [{\citenamefont {Thompson}(2010)}]{2010THO_ar}%
  \BibitemOpen
  \bibfield  {author} {\bibinfo {author} {\bibfnamefont {M.~B.}\ \bibnamefont
  {Thompson}}\ }(\bibinfo {year} {2010})\BibitemShut {NoStop}%
\bibitem [{\citenamefont {Fishman}(1978)}]{1978FIS}%
  \BibitemOpen
  \bibfield  {author} {\bibinfo {author} {\bibfnamefont {G.}~\bibnamefont
  {Fishman}},\ }\href@noop {} {\  (\bibinfo {year} {1978})}\BibitemShut
  {NoStop}%
\bibitem [{\citenamefont {Yousefi}(2011)}]{2011YOU}%
  \BibitemOpen
  \bibfield  {author} {\bibinfo {author} {\bibfnamefont {S.}~\bibnamefont
  {Yousefi}},\ }\emph {\bibinfo {title} {MSER-5Y: An Improved Version of MSER-5
  with Automatic Confidence Interval Estimation.}},\ \href@noop {} {Ph.D.
  thesis},\ \bibinfo  {school} {North Carolina State University} (\bibinfo
  {year} {2011})\BibitemShut {NoStop}%
\bibitem [{\citenamefont {Franklin}\ and\ \citenamefont
  {White}(2008)}]{2008FRA}%
  \BibitemOpen
  \bibfield  {author} {\bibinfo {author} {\bibfnamefont {W.~W.}\ \bibnamefont
  {Franklin}}\ and\ \bibinfo {author} {\bibfnamefont {K.~P.}\ \bibnamefont
  {White}},\ }in\ \href {\doibase 10.1109/WSC.2008.4736111} {\emph {\bibinfo
  {booktitle} {2008 Winter Simulation Conference}}}\ (\bibinfo {year} {2008})\
  pp.\ \bibinfo {pages} {541--546}\BibitemShut {NoStop}%
\bibitem [{\citenamefont {Spencer}\ \emph {et~al.}(2014)\citenamefont
  {Spencer}, \citenamefont {Blunt}, \citenamefont {Vigor}, \citenamefont
  {Malone}, \citenamefont {Foulkes}, \citenamefont {Shepherd},\ and\
  \citenamefont {Thom}}]{2014SPE}%
  \BibitemOpen
  \bibfield  {author} {\bibinfo {author} {\bibfnamefont {J.~S.}\ \bibnamefont
  {Spencer}}, \bibinfo {author} {\bibfnamefont {N.~S.}\ \bibnamefont {Blunt}},
  \bibinfo {author} {\bibfnamefont {W.~A.}\ \bibnamefont {Vigor}}, \bibinfo
  {author} {\bibfnamefont {F.~D.}\ \bibnamefont {Malone}}, \bibinfo {author}
  {\bibfnamefont {W.~M.~C.}\ \bibnamefont {Foulkes}}, \bibinfo {author}
  {\bibfnamefont {J.~J.}\ \bibnamefont {Shepherd}}, \ and\ \bibinfo {author}
  {\bibfnamefont {A.~J.~W.}\ \bibnamefont {Thom}}\ }(\bibinfo {year}
  {2014})\BibitemShut {NoStop}%
\bibitem [{\citenamefont {Spencer}\ \emph {et~al.}(2019)\citenamefont
  {Spencer}, \citenamefont {Blunt}, \citenamefont {Choi}, \citenamefont
  {Etrych}, \citenamefont {Filip}, \citenamefont {Foulkes}, \citenamefont
  {Franklin}, \citenamefont {Handley}, \citenamefont {Malone}, \citenamefont
  {Neufeld}, \citenamefont {Di~Remigio}, \citenamefont {Rogers}, \citenamefont
  {Scott}, \citenamefont {Shepherd}, \citenamefont {Vigor}, \citenamefont
  {Weston}, \citenamefont {Xu},\ and\ \citenamefont {Thom}}]{2019SPE}%
  \BibitemOpen
  \bibfield  {author} {\bibinfo {author} {\bibfnamefont {J.~S.}\ \bibnamefont
  {Spencer}}, \bibinfo {author} {\bibfnamefont {N.~S.}\ \bibnamefont {Blunt}},
  \bibinfo {author} {\bibfnamefont {S.}~\bibnamefont {Choi}}, \bibinfo {author}
  {\bibfnamefont {J.}~\bibnamefont {Etrych}}, \bibinfo {author} {\bibfnamefont
  {M.-A.}\ \bibnamefont {Filip}}, \bibinfo {author} {\bibfnamefont {W.~M.~C.}\
  \bibnamefont {Foulkes}}, \bibinfo {author} {\bibfnamefont {R.~S.~T.}\
  \bibnamefont {Franklin}}, \bibinfo {author} {\bibfnamefont {W.~J.}\
  \bibnamefont {Handley}}, \bibinfo {author} {\bibfnamefont {F.~D.}\
  \bibnamefont {Malone}}, \bibinfo {author} {\bibfnamefont {V.~A.}\
  \bibnamefont {Neufeld}}, \bibinfo {author} {\bibfnamefont {R.}~\bibnamefont
  {Di~Remigio}}, \bibinfo {author} {\bibfnamefont {T.~W.}\ \bibnamefont
  {Rogers}}, \bibinfo {author} {\bibfnamefont {C.~J.~C.}\ \bibnamefont
  {Scott}}, \bibinfo {author} {\bibfnamefont {J.~J.}\ \bibnamefont {Shepherd}},
  \bibinfo {author} {\bibfnamefont {W.~A.}\ \bibnamefont {Vigor}}, \bibinfo
  {author} {\bibfnamefont {J.}~\bibnamefont {Weston}}, \bibinfo {author}
  {\bibfnamefont {R.}~\bibnamefont {Xu}}, \ and\ \bibinfo {author}
  {\bibfnamefont {A.~J.}\ \bibnamefont {Thom}},\ }\href {\doibase
  10.1021/acs.jctc.8b01217} {\bibfield  {journal} {\bibinfo  {journal} {Journal
  of Chemical Theory and Computation}\ }\textbf {\bibinfo {volume} {15}},\
  \bibinfo {pages} {1728} (\bibinfo {year} {2019})},\ \bibinfo {note} {pMID:
  30681844},\ \Eprint
  {http://arxiv.org/abs/https://doi.org/10.1021/acs.jctc.8b01217}
  {https://doi.org/10.1021/acs.jctc.8b01217} \BibitemShut {NoStop}%
\bibitem [{Note()}]{Note}%
  \BibitemOpen
  \bibinfo {note} {The correlation length $\tau $ is differently defined in two
  papers, \cite {1986STR, 2010THO_ar} but they give the same definition of
  error. We took the newer definition of $\tau $.\cite
  {2010THO_ar}}\BibitemShut {NoStop}%
\bibitem [{\citenamefont {Wei}(2006)}]{2006WEI}%
  \BibitemOpen
  \bibfield  {author} {\bibinfo {author} {\bibfnamefont {W.}~\bibnamefont
  {Wei}},\ }\href@noop {} {\emph {\bibinfo {title} {Time Series Analysis:
  Univariate and Multivariate Methods}}}\ (\bibinfo {year} {2006})\BibitemShut
  {NoStop}%
\bibitem [{\citenamefont {Akaike}(1992)}]{1992AKA}%
  \BibitemOpen
  \bibfield  {author} {\bibinfo {author} {\bibfnamefont {H.}~\bibnamefont
  {Akaike}},\ }\enquote {\bibinfo {title} {Information theory and an extension
  of the maximum likelihood principle},}\ in\ \href {\doibase
  10.1007/978-1-4612-0919-5_38} {\emph {\bibinfo {booktitle} {Breakthroughs in
  Statistics: Foundations and Basic Theory}}},\ \bibinfo {editor} {edited by\
  \bibinfo {editor} {\bibfnamefont {S.}~\bibnamefont {Kotz}}\ and\ \bibinfo
  {editor} {\bibfnamefont {N.~L.}\ \bibnamefont {Johnson}}}\ (\bibinfo
  {publisher} {Springer New York},\ \bibinfo {address} {New York, NY},\
  \bibinfo {year} {1992})\ pp.\ \bibinfo {pages} {610--624}\BibitemShut
  {NoStop}%
\bibitem [{\citenamefont {Flyvbjerg}\ and\ \citenamefont
  {Petersen}(1989)}]{1989FLY}%
  \BibitemOpen
  \bibfield  {author} {\bibinfo {author} {\bibfnamefont {H.}~\bibnamefont
  {Flyvbjerg}}\ and\ \bibinfo {author} {\bibfnamefont {H.~G.}\ \bibnamefont
  {Petersen}},\ }\href {\doibase 10.1063/1.457480} {\bibfield  {journal}
  {\bibinfo  {journal} {The Journal of Chemical Physics}\ }\textbf {\bibinfo
  {volume} {91}},\ \bibinfo {pages} {461} (\bibinfo {year} {1989})},\ \Eprint
  {http://arxiv.org/abs/https://doi.org/10.1063/1.457480}
  {https://doi.org/10.1063/1.457480} \BibitemShut {NoStop}%
\bibitem [{\citenamefont {Needs}\ \emph {et~al.}(2015)\citenamefont {Needs},
  \citenamefont {Towler}, \citenamefont {Drummond},\ and\ \citenamefont
  {R\'{\i}os}}]{casino_manual}%
  \BibitemOpen
  \bibfield  {author} {\bibinfo {author} {\bibfnamefont {R.}~\bibnamefont
  {Needs}}, \bibinfo {author} {\bibfnamefont {M.}~\bibnamefont {Towler}},
  \bibinfo {author} {\bibfnamefont {N.}~\bibnamefont {Drummond}}, \ and\
  \bibinfo {author} {\bibfnamefont {P.~L.}\ \bibnamefont {R\'{\i}os}},\ }\href
  {https://vallico.net/casinoqmc/} {\emph {\bibinfo {title} {Casino 2.13 User
  Manual}}}\ (\bibinfo  {publisher} {Cambridge University Press, Cambridge},\
  \bibinfo {year} {2015})\BibitemShut {NoStop}%
\bibitem [{\citenamefont {Hehre}\ \emph {et~al.}(1969)\citenamefont {Hehre},
  \citenamefont {Stewart},\ and\ \citenamefont {Pople}}]{1969HEH}%
  \BibitemOpen
  \bibfield  {author} {\bibinfo {author} {\bibfnamefont {W.~J.}\ \bibnamefont
  {Hehre}}, \bibinfo {author} {\bibfnamefont {R.~F.}\ \bibnamefont {Stewart}},
  \ and\ \bibinfo {author} {\bibfnamefont {J.~A.}\ \bibnamefont {Pople}},\
  }\href {\doibase 10.1063/1.1672392} {\bibfield  {journal} {\bibinfo
  {journal} {The Journal of Chemical Physics}\ }\textbf {\bibinfo {volume}
  {51}},\ \bibinfo {pages} {2657} (\bibinfo {year} {1969})},\ \Eprint
  {http://arxiv.org/abs/https://doi.org/10.1063/1.1672392}
  {https://doi.org/10.1063/1.1672392} \BibitemShut {NoStop}%
\bibitem [{\citenamefont {Ma}\ \emph {et~al.}(2005)\citenamefont {Ma},
  \citenamefont {Towler}, \citenamefont {Drummond},\ and\ \citenamefont
  {Needs}}]{2005MA}%
  \BibitemOpen
  \bibfield  {author} {\bibinfo {author} {\bibfnamefont {A.}~\bibnamefont
  {Ma}}, \bibinfo {author} {\bibfnamefont {M.~D.}\ \bibnamefont {Towler}},
  \bibinfo {author} {\bibfnamefont {N.~D.}\ \bibnamefont {Drummond}}, \ and\
  \bibinfo {author} {\bibfnamefont {R.~J.}\ \bibnamefont {Needs}},\ }\href
  {\doibase 10.1063/1.1940588} {\bibfield  {journal} {\bibinfo  {journal} {The
  Journal of Chemical Physics}\ }\textbf {\bibinfo {volume} {122}},\ \bibinfo
  {pages} {224322} (\bibinfo {year} {2005})},\ \Eprint
  {http://arxiv.org/abs/https://doi.org/10.1063/1.1940588}
  {https://doi.org/10.1063/1.1940588} \BibitemShut {NoStop}%
\bibitem [{\citenamefont {Kato}(1957)}]{1957KAT}%
  \BibitemOpen
  \bibfield  {author} {\bibinfo {author} {\bibfnamefont {T.}~\bibnamefont
  {Kato}},\ }\href {\doibase 10.1002/cpa.3160100201} {\bibfield  {journal}
  {\bibinfo  {journal} {Communications on Pure and Applied Mathematics}\
  }\textbf {\bibinfo {volume} {10}},\ \bibinfo {pages} {151} (\bibinfo {year}
  {1957})},\ \Eprint
  {http://arxiv.org/abs/https://onlinelibrary.wiley.com/doi/pdf/10.1002/cpa.3160100201}
  {https://onlinelibrary.wiley.com/doi/pdf/10.1002/cpa.3160100201} \BibitemShut
  {NoStop}%
\bibitem [{\citenamefont {Dunning}(1989)}]{1989DUN}%
  \BibitemOpen
  \bibfield  {author} {\bibinfo {author} {\bibfnamefont {T.~H.}\ \bibnamefont
  {Dunning}},\ }\href {\doibase 10.1063/1.456153} {\bibfield  {journal}
  {\bibinfo  {journal} {The Journal of Chemical Physics}\ }\textbf {\bibinfo
  {volume} {90}},\ \bibinfo {pages} {1007} (\bibinfo {year} {1989})},\ \Eprint
  {http://arxiv.org/abs/https://doi.org/10.1063/1.456153}
  {https://doi.org/10.1063/1.456153} \BibitemShut {NoStop}%
\bibitem [{\citenamefont {Parrish}\ \emph {et~al.}(2017)\citenamefont
  {Parrish}, \citenamefont {Burns}, \citenamefont {Smith}, \citenamefont
  {Simmonett}, \citenamefont {DePrince}, \citenamefont {Hohenstein},
  \citenamefont {Bozkaya}, \citenamefont {Sokolov}, \citenamefont {Di~Remigio},
  \citenamefont {Richard}, \citenamefont {Gonthier}, \citenamefont {James},
  \citenamefont {McAlexander}, \citenamefont {Kumar}, \citenamefont {Saitow},
  \citenamefont {Wang}, \citenamefont {Pritchard}, \citenamefont {Verma},
  \citenamefont {Schaefer}, \citenamefont {Patkowski}, \citenamefont {King},
  \citenamefont {Valeev}, \citenamefont {Evangelista}, \citenamefont {Turney},
  \citenamefont {Crawford},\ and\ \citenamefont {Sherrill}}]{2017PAR}%
  \BibitemOpen
  \bibfield  {author} {\bibinfo {author} {\bibfnamefont {R.~M.}\ \bibnamefont
  {Parrish}}, \bibinfo {author} {\bibfnamefont {L.~A.}\ \bibnamefont {Burns}},
  \bibinfo {author} {\bibfnamefont {D.~G.~A.}\ \bibnamefont {Smith}}, \bibinfo
  {author} {\bibfnamefont {A.~C.}\ \bibnamefont {Simmonett}}, \bibinfo {author}
  {\bibfnamefont {A.~E.}\ \bibnamefont {DePrince}}, \bibinfo {author}
  {\bibfnamefont {E.~G.}\ \bibnamefont {Hohenstein}}, \bibinfo {author}
  {\bibfnamefont {U.~u.}\ \bibnamefont {Bozkaya}}, \bibinfo {author}
  {\bibfnamefont {A.~Y.}\ \bibnamefont {Sokolov}}, \bibinfo {author}
  {\bibfnamefont {R.}~\bibnamefont {Di~Remigio}}, \bibinfo {author}
  {\bibfnamefont {R.~M.}\ \bibnamefont {Richard}}, \bibinfo {author}
  {\bibfnamefont {J.~r. m.~F.}\ \bibnamefont {Gonthier}}, \bibinfo {author}
  {\bibfnamefont {A.~M.}\ \bibnamefont {James}}, \bibinfo {author}
  {\bibfnamefont {H.~R.}\ \bibnamefont {McAlexander}}, \bibinfo {author}
  {\bibfnamefont {A.}~\bibnamefont {Kumar}}, \bibinfo {author} {\bibfnamefont
  {M.}~\bibnamefont {Saitow}}, \bibinfo {author} {\bibfnamefont
  {X.}~\bibnamefont {Wang}}, \bibinfo {author} {\bibfnamefont {B.~P.}\
  \bibnamefont {Pritchard}}, \bibinfo {author} {\bibfnamefont {P.}~\bibnamefont
  {Verma}}, \bibinfo {author} {\bibfnamefont {H.~F.}\ \bibnamefont {Schaefer}},
  \bibinfo {author} {\bibfnamefont {K.}~\bibnamefont {Patkowski}}, \bibinfo
  {author} {\bibfnamefont {R.~A.}\ \bibnamefont {King}}, \bibinfo {author}
  {\bibfnamefont {E.~F.}\ \bibnamefont {Valeev}}, \bibinfo {author}
  {\bibfnamefont {F.~A.}\ \bibnamefont {Evangelista}}, \bibinfo {author}
  {\bibfnamefont {J.~M.}\ \bibnamefont {Turney}}, \bibinfo {author}
  {\bibfnamefont {T.~D.}\ \bibnamefont {Crawford}}, \ and\ \bibinfo {author}
  {\bibfnamefont {C.~D.}\ \bibnamefont {Sherrill}},\ }\href {\doibase
  10.1021/acs.jctc.7b00174} {\bibfield  {journal} {\bibinfo  {journal} {Journal
  of Chemical Theory and Computation}\ }\textbf {\bibinfo {volume} {13}},\
  \bibinfo {pages} {3185} (\bibinfo {year} {2017})},\ \bibinfo {note} {pMID:
  28489372},\ \Eprint
  {http://arxiv.org/abs/https://doi.org/10.1021/acs.jctc.7b00174}
  {https://doi.org/10.1021/acs.jctc.7b00174} \BibitemShut {NoStop}%
\end{thebibliography}%
\end{document}